 \documentclass[12pt]{article}
 \usepackage[cp1251]{inputenc} 
 \usepackage[english, russian]{babel} 
 \usepackage{amsmath, latexsym, amssymb, bm, array, graphics, eucal,
 amsfonts,}

 \renewcommand{\baselinestretch}{1.2}
\usepackage[dvips]{graphicx}

\begin{document}
\thispagestyle{empty}
\large
\renewcommand{\abstractname}{Abstract}
\renewcommand{\refname}{\begin{center}
 REFERENCES\end{center}}
\newcommand{\mc}[1]{\mathcal{#1}}
\newcommand{\E}{\mc{E}}
\makeatother

\begin{center}
\bf Intrinsic nonlinearity of interaction of an
electromagnetic field with quantum plasma and its research
\end{center} \medskip
\begin{center}
  \bf A. V. Latyshev\footnote{$avlatyshev@mail.ru$} and
  A. A. Yushkanov\footnote{$yushkanov@inbox.ru$}
\end{center}\medskip

\begin{center}
{\it Faculty of Physics and Mathematics,\\ Moscow State Regional
University, 105005,\\ Moscow, Radio str., 10A}
\end{center}\medskip

\begin{abstract}
The analysis of nonlinear interaction of transversal
electromagnetic field with quantum collisionless plasma is carried out.
Formulas for calculation electric current in quantum collisionless plasma
at any temperature are deduced.
It has appeared, that the nonlinearity account leads to occurrence
of the longitudinal electric current directed along a wave vector.
This second current is orthogonal to the known transversal classical current,
received at the classical linear analysis.
The case of degenerate electronic plasma is considered.
The concept of longitudinal-transversal conductivity is entered.
The graphic analysis of the real and imaginary parts of dimensionless
coefficient of longitudinal-transversal conductivity is made.
It is shown, that for degenerate plasmas the electric current is
calculated under the formula, not containing quadratures.
In this formula we have allocated known Kohn's singularities (W. Kohn, 1959).


\medskip

electron gas
\end{abstract}

\begin{center}
\bf  Introduction
\end{center}
Dielectric permeability in quantum plasma was studied by many
aut\-hors \cite{Klim} -- \cite{Dressel}.
Dielectric permeability is one of the major plasma charac\-te\-ris\-tics.

This quantity is necessary for the description of skin-effect
\cite{Gelder}, for the analysis surface plasmons \cite{Fuchs},
for descriptions of process of propagation and attenuation of the
transversal plasma oscillations \cite{Shukla2}, for studying of
the mechanism of penetration electromagnetic waves in plasma
\cite{Shukla1}, and for the analysis of other problems in the
plasma physics \cite{Brod} -- \cite {Lat4}.

Let us notice, that for the first time in work \cite{Klim} the formula
for cal\-cu\-la\-tion of longitudinal dielectric permeability into
quantum plasma has been deduced. Then the same formula has been deduced
and in work \cite{Lin}.

In the present work formulas for calculation electric current
into quantum collisionless plasma  at any temperature (at any
degrees of degeneration of the electronic gas)  are deduced.

Here the approach developed by Klimontovich Silin \cite{Klim} is
generalized.

At the solution of Schr\"odinger equation we consider and in expansion
of distribution Wigner  function, and in Wigner---Vlasov integral
expan\-sion the quantities proportional
to square of potential of an external electro\-mag\-ne\-tic field.

It has appeared, that electric current expression consists of two sum\-mands.
The first summand, linear on vector potential, is
known classical expression of an electric current.
This electric current is directed along vector potential electromagnetic
field. The second summand represents itself an electric current,
which is proportional to the square
vector potential of electromagnetic fields. The second current
it is perpendicular to the first and it is directed along the  wave
vector.
Occurrence of the second current comes to light the spent account
intrinsic nonlinear character
interactions of an electromagnetic field with quantum plasma.

For the case of degenerate quantum plasma expression of the electric
current, not containing quadratures, is received.
At the deducing of this expression
Landau' rule for calculation singular integrals is used.
At use of this rule calculation
these integrals containing a pole on the real axis,
it is carried out by means of integration on infinitesimal
half-circles in the bottom half-plane with the centre in this pole.

\begin{center}
  {\bf 1. Kinetic equation for Wigner function}
\end{center}

Let us consider Shr\"odinger equation which has been written down
for a particle in an electromagnetic field on a density matrix
$\rho$
$$
i\hbar \dfrac{\partial \rho}{\partial t}=H\rho-{H^*}'\rho.
$$

Here $H$ is the Hamilton operator, $H^*$ is the compex conjugate operator to
$H$, ${H^*}'$ is the compex conjugate operator to
$H$, which operates on the shaded spatial variables
$\mathbf{r}'$.
We believe that the scalar potential is equal to zero.

In work \cite{Lat1} it is shown that the Shr\"odinger equation
under condition of calibration of potential of electromagnetic
field
$$
\nabla\cdot {\bf A}=0
$$
will be transformed to the kinetic equation
$$
\dfrac{\partial f}{\partial t}+
\mathbf{v}\dfrac{\partial f}{\partial {\bf r}}=W[f],
\eqno{(1.1)}
$$
written down in regard to Wigner function
$$
f(\mathbf{r},\mathbf{p},t)=\int
\rho(\mathbf{r}+\dfrac{\mathbf{a}}{2},\mathbf{r}-
\dfrac{\mathbf{a}}{2},t)e^{-i\mathbf{p}\mathbf{a}/\hbar}d^3a,
$$
besides
$$
\rho(\mathbf{R},\mathbf{R}',t)=\dfrac{1}{(2\pi \hbar)^3}
\int f(\dfrac{\mathbf{R}+\mathbf{R}'}{2}, \mathbf{p},t)
e^{i\mathbf{p}(\mathbf{R}-\mathbf{R}')/\hbar}d^3p.
$$

Here $e$ is the electron charge, $m$ is the electron mass, $c$ is the
speed of light.

Wigner integral \cite{Lat1} equals
$$
W[f]=
\iint\Bigg\{
\dfrac{e}{2mc}
\Big[\mathbf{A}(\mathbf{r}+\dfrac{\mathbf{a}}{2},t)+
\mathbf{A}(\mathbf{r}-\dfrac{\mathbf{a}}{2},t)-
2\mathbf{A}(\mathbf{r},t)\Big]\nabla f+
$$\smallskip
$$
+\dfrac{ie}{ mc\hbar}\Big[
\mathbf{A}(\mathbf{r}+\dfrac{\mathbf{a}}{2},t)
-\mathbf{A}(\mathbf{r}-\dfrac{\mathbf{a}}{2},t)\Big]\mathbf{p'}
f+
$$\smallskip
$$-
\dfrac{i e^2}{2 mc^2\hbar}\Big[\mathbf{A}^2(\mathbf{r}+
\dfrac{\mathbf{a}}{2},t)-
\mathbf{A}^2(\mathbf{r}-\dfrac{\mathbf{a}}{2},t)\Big]f\Bigg\}
e^{i(\mathbf{p'}-\mathbf{p})\mathbf{a}/\hbar}
\dfrac{d^3a\,d^3p'}{(2\pi\hbar)^3}.
$$

Vector potential of an electromagnetic field we take orthogonal
to direction of a wave vector $ {\bf k} $ (${\bf k \cdot A} =0$) in the form
running harmonious wave
${\bf A}({\bf r},t)={\bf A}_0e^{i({\bf kr}-\omega t)}$.

We denote further
$$
f_{\pm}\equiv f({\bf r,p}\mp \hbar k/2,t)\qquad
f_{\pm\pm}\equiv f({\bf r,p}\mp \hbar k,t).
$$

Now we transform the Wigner integral to the following form
$$
W[f]=\dfrac{e {\bf A}}{2mc}\Big(\dfrac{\partial f_+}{\partial {\bf r}}+
\dfrac{\partial f_-}{\partial{\bf r}}-2\dfrac{\partial f}
{\partial {\bf r}}\Big)+\dfrac{ie {\bf A}}{mc\hbar}{\bf p}(f_+-f_-)-
$$
$$
-\dfrac{ie^2{\bf A}}{2mc^2\hbar}(f_{++}-f_{--}).
\eqno{(1.2)}
$$

We will enter the locally equalibrium and absolute Fermi---Dirac
distribution $f^{(0)}$ and $f_F$,
$$
f^{(0)}=f^{(0)}({\bf P},\mathbf{r},t)=[1+\exp(C^2-\alpha)]^{-1},
$$
and
$$
f_F=f_F(P)=[1+\exp(P^2-\alpha)]^{-1},
$$

Here
$$
{\bf C}\equiv\mathbf{C}({\bf P},\mathbf{r},t)=\dfrac{\mathbf{v}}{v_T}=
{\bf P}-\dfrac{e}{cp_T}\mathbf{A}(\mathbf{r},t),
\qquad\qquad \alpha=\dfrac{\mu}{k_BT},
$$
$\mathbf{C}$ is the dimensionless electron velocity,
$v_T={1}/{\sqrt{\beta}}$ is the thermal electron velocity,
$\beta={m}/{2k_BT}$, ${\bf P}={{\bf p}}/{p_T}$ is the dimensionless
electron momentum, $m$ is the electron mass, $k_B$
is the Boltzmann constant,
$T$ is the plasma temperature, $\mu$ is the chemical potential
electronic gas, $\alpha$ is the dimensionless chemical potential.

Let us operate with a method consecutive approximations.
In square-law approach on vector potential $ {\bf A} $
$f $ in the first summand in (1.2) it is necessary to replace
Wigner function on locally equilibrium distribution $f^{(0)} $, in the third
 summand --- on absolute distribution $f_F $, and in second ---
on its linear approach found in \cite{Lat1}, i.e.
to put $f=f^{(1)} $, where
$$
f^{(1)}=f^{(0)}-{\bf PA}\Big(\dfrac{2e}{cp_T}g(P)+
\dfrac{ev_T}{c\hbar}\dfrac{f_F^+-f_F^-}{\omega-v_T{\bf kP}}\Big),
\eqno{(1.3)}
$$
where
$$
g(P)=e^{P^2-\alpha}\Big(1+e^{P^2-\alpha}\Big)^{-2},\qquad
f_F^{\pm}=[1+e^{P^2_{\pm}-\alpha}]^{-1},
$$
$$
P^2_{\pm}=\Big({\bf P}\mp\dfrac{\hbar {\bf k}}{2p_T}\Big)^2.
$$

Let us notice, that in linear approximation
$$
f^{(0)}=f_F+{\bf PA}\dfrac{2e}{cp_T}g(P).
$$
Hence, function $f^{(1)} $ is represented in the form
$$
f^{(1)}=f_F(P)-\dfrac{ev_T}{c\hbar}\mathbf{PA}(\mathbf{r},t)
\dfrac{f_F^+-f_F^-}{\omega-v_T\mathbf{kP}}.
$$

Let us show, that the first  summand in Wigner integral (1.2)
equally to zero. According to problem statement vector potential
of electromagnetic field varies along an axis $x $. Hence,
gradient of locally equilibrium distribution of Fermi---Dirac
is proportional to the vector $\mathbf{k}$:
$\partial f^{(0)}_{\pm}/\partial {\bf r}\sim \mathbf{k}$,
$\partial f^{(0)}/\partial {\bf r}\sim \mathbf{k}$.
Therefore
$$
\mathbf{A}\Big[\dfrac{\partial f^{(0)}_+}{\partial {\bf r}}+
\dfrac{\partial f^{(0)}_-}{\partial {\bf  r}}-
2\dfrac{\partial f^{(0)}}{\partial {\bf r}}\Big]\sim\mathbf{Ak}=0.
$$

We notice that
$$
f_F^{+-}=f_F^{-+}=f_F=f_F(P).
$$
Therefore, Wigner integral (1.2) equals
$$
W[f]=\dfrac{iev_T}{c\hbar}\mathbf{P}\mathbf{A}
\Big[f_F^+-f_F^--\dfrac{ev_T}{c\hbar}\mathbf{PA}(\mathbf{r},t)
\dfrac{f_F^{++}+f_F^{--}-2f_F}{\omega-v_T\mathbf{kP}}\Big]-
$$
$$
-\mathbf{A}^2\dfrac{i e^2}{2 mc^2\hbar}
\Big(f_F^{++}-f_F^{--}\Big).
\eqno{(1.4)}
$$\medskip

Here
$$
f_F^{\pm\pm}=\Big[1+e^{P^2_{\pm\pm}-\alpha}\Big]^{-1},\qquad
P^2_{\pm\pm}=\Big({\bf P}\mp\dfrac{\hbar {\bf k}}{p_T}\Big)^2.
$$

Let us return to the decision of the equation (1.1) with
Wigner integral (1.4).
Let us search for Wigner function in the form,
square-law concerning of vector potential
${\bf A}=\mathbf{A}(\mathbf{r},t)$:
$$
f=f_F(P)-\dfrac{ev_T}{c\hbar}\mathbf{PA}
\dfrac{f_F^+-f_F^-}{\omega-v_T\mathbf{kP}}+\mathbf{A}^2
h(\mathbf{P}),
$$
where $h(\mathbf{P})$ is the new unknown function.

We receive the equation from which it is found
$$
\mathbf{A}^2h(\mathbf{P})=
\dfrac{(ev_T)^2}{2(c\hbar)^2}[\mathbf{PA}]^2
\dfrac{f_F^{++}+f_F^{--}-2f_F}{(\omega-v_T\mathbf{kP})^2}+
\dfrac{e^2}{4mc^2\hbar}\mathbf{A}^2\dfrac{f_F^{++}-
f_F^{--}}{\omega-v_T\mathbf{kP}}.
$$

By means of last two equalities let us construct the
Wigner function Вигнера in the second approximation on
vector potential
$\mathbf{A}(\mathbf{r},t)$:
$$
f=f^{(0)}-\mathbf{PA}(\mathbf{r},t)\Big[\dfrac{2e}{cp_T}g(P)+
\dfrac{ev_T}{c\hbar}\dfrac{f_F^+-f_F^-}{\omega-v_T\mathbf{kP}}\Big]+
$$
$$
+\dfrac{(ev_T)^2}{2(c\hbar)^2}[\mathbf{PA}]^2
\dfrac{f_F^{++}+f_F^{--}-2f_F}{(\omega-v_T\mathbf{kP})^2}+
\dfrac{e^2}{4mc^2\hbar}\mathbf{A}^2\dfrac{f_F^{++}-
f_F^{--}}{\omega-v_T\mathbf{kP}}.
\eqno{(1.5)}
$$ \medskip

This function represents square-law decomposition of
distribution function on vector potential
$\mathbf{A}(\mathbf{r},t)$.

\begin{center}
\bf  2. Density of electric current in quantum plasmas
\end{center}

By definition, the density of electric current is equal
$$
\mathbf{j}(\mathbf{r},t)=e\int \mathbf{v}(\mathbf{r},\mathbf{p},t)
f(\mathbf{r},\mathbf{p},t)\dfrac{2\,d^3p}{(2\pi\hbar)^3}.
\eqno{(2.1)}
$$

Substituting in equality (2.1) explicit expression for velocity
$$
\mathbf{v}(\mathbf{r},\mathbf{P},t)=
\dfrac{\mathbf{p}}{m}-\dfrac{e \mathbf{A}(\mathbf{r},t)}{mc}=
\dfrac{p_T\mathbf{P}}{m}-\dfrac{e \mathbf{A}(\mathbf{r},t)}{mc}=
v_T\mathbf{P}-\dfrac{e \mathbf{A}(\mathbf{r},t)}{mc}.
$$
and, leaving linear and quadratic (square-law) expressions concerning
vector potential of the field, we receive \medskip
$$
\mathbf{j}(\mathbf{r},t)=-
\dfrac{2e^2p_T^4}{(2\pi \hbar)^3mc}\int
\mathbf{P}\big[\mathbf{P}\mathbf{A}\big]\Big[
\dfrac{2}{p_T}g(P)+\dfrac{v_T}{\hbar}\dfrac{f_F^+-f_F^-}{\omega
-v_T\mathbf{kP}}\Big]d^3P+
$$ \medskip
$$
+\dfrac{2e^3p_T^3}{(2\pi \hbar)^3mc^2}\mathbf{A}
\int\big[\mathbf{P}\mathbf{A}\big]
\Big[
\dfrac{2}{p_T}g(P)+\dfrac{v_T}{\hbar}\dfrac{f_F^+-f_F^-}{\omega
-v_T\mathbf{kP}}\Big]d^3P+
$$ \medskip
$$
+\dfrac{2e^3p_T^4}{(2\pi\hbar)^3mc^2\hbar}
\int\mathbf{P} \Bigg[\dfrac{mv_T^2}{2\hbar}[\mathbf{PA}]^2
\dfrac{f_F^{++}+f_F^{--}-2f_F}
{(\omega-v_T\mathbf{kP})^2}+\dfrac{\mathbf{A}^2}{4}
\dfrac{f_F^{++}-f_F^{--}}{\omega-v_T\mathbf{kP}}\Bigg]d^3P.
\eqno{(2.2)}
$$ \medskip

The first summand in (2.2) is linear expression
of the density of electric current,
$$
\mathbf{j}_{\rm linear}(\mathbf{r},t)=$$$$=-
\dfrac{2e^2p_T^4}{(2\pi \hbar)^3mc}\int
\mathbf{P}\big[\mathbf{P}\mathbf{A}\big]\Big[
\dfrac{2}{p_T}g(P)+\dfrac{v_T}{\hbar}\dfrac{f_F^+-f_F^-}{\omega
-v_T\mathbf{kP}}\Big]d^3P.
\eqno{(2.3)}
$$
found, in particular, in our previous
work \cite{Lat1}.
The second summand is the square-law amendment
to the first. The third summand represents density of
{\it longitudinal} electric current, unlike density
of classical {\it trans\-ver\-sal} electric current described
first two summands.

Thus, in square-law approximation on vector potential
of elec\-tro\-mag\-ne\-tic field it has appeared, that vector potential
electromagnetic fields generates also the longitudinal electric current
besides the transversal  current (2.3).

Vector potential of the field we will direct along an axis $y $:
${\bf A}$=$A(x,t)$\\$\{0,1,0\}$, $A(x,t)=A_ye^{i(kx-\omega t)}$,
and wave vector ${\bf k}$ we direct along axis $x$:
${\bf k}=k\{1,0,0\}$. According to (2.2) the longitudinal equals
${\bf j}_{\rm long}=j_{\rm long}(x,t)\{1,0,0\}$, where
$$
j_{\rm long}(x,t)=
\dfrac{e^3p_T^2A^2(x,t)}{(2\pi\hbar)^3mc^2q^2}\times
$$
$$
\times \Bigg[\int
\dfrac{f_F^{++}+f_F^{--}-2f_F}{(P_x-\omega/v_Tk)^2}P_xP_y^2d^3P-
\dfrac{q}{2}\int\dfrac{f_F^{++}-f_F^{--}}{P_x-\omega/v_Tk}P_xd^3P\Bigg].
\eqno{(2.4)}
$$

Here $q=k/k_T, k_T=mv_T/\hbar$.

Let us simplify the formula (2.4), having calculated internal integrals in
planes $ (P_y, P_z) $.  We receive as the result, that
$$
j_{\rm long}(x,t)=
\dfrac{e^3p_T^2A^2(x,t)}{(2\pi\hbar)^3mc^2q^2}\times
$$
$$
\times \Bigg[\int\limits_{\infty}^{\infty}
\dfrac{P_xL(P_x,\alpha)dP_x}{(P_x-\omega/v_Tk)^2}-
\dfrac{q}{2}\int\limits_{-\infty}^{\infty}
\ln\dfrac{1+e^{-{P_x^+}^2+\alpha}}{1+e^{-{P_x^-}^2+\alpha}}
\dfrac{dP_x}{P_x-\omega/v_Tk}\Bigg],
$$
where
$$
L(P_x,\alpha)=\int\limits_{0}^{\infty}\dfrac{(1+e^{-{P_x^{++}}^2+\alpha})^\rho
(1+e^{-{P_x^{--}}^2+\alpha})^\rho}{(1+e^{-{P_x}^2+\alpha})^\rho}d\rho,
$$
$$
{P_x^{\pm}}=P_x\mp\dfrac{\hbar k}{2p_T},\qquad
P_x^{\pm\pm}=P_x\mp\dfrac{\hbar k}{p_T}.
$$

At calculation "dispersing"\, integrals it is necessary to use
known Landau rule, bypassing a pole on the real axis on
the half-circle laying in the bottom half-plane, preliminary
having executed integration in parts. It is equivalent to the following
application of the formula Sokhotsky
$$
\int\limits_{a}^{b}\dfrac{\varphi(\tau)d\tau}{(\tau-x)^2}=
\lim\limits_{\varepsilon\to 0}\int\limits_{a}^{b}\dfrac{\varphi(\tau)d\tau}
{[\tau-(x+i\varepsilon)]^2}=
\lim\limits_{\varepsilon\to 0}
\Big[-\dfrac{\varphi(\tau)}{\tau-(x+i\varepsilon)}\Bigg|_{a}^{b}+
$$
$$
+\int\limits_{a}^{b}\dfrac{\varphi'(\tau)d\tau}{\tau-(x+i\varepsilon)}\Big]=
-\dfrac{\varphi(\tau)}{\tau-x}\Bigg|_a^b+i\pi \varphi'(x)+
\int\limits_{a}^{b}\dfrac{\varphi'(\tau)d\tau}{\tau-x}.
\eqno{(2.5)}
$$

\begin{center}
  \bf 3. Degenerate plasmas
\end{center}

Let's consider the case of degenerate plasmas.

In the formula (2.4) we will pass to the limit at $T\to 0$ and
we will carry out replacement of one
variable of integration
$$
P_x\to \dfrac{v_F}{v_T}P_x,
$$
where $v_F$ is the electron velocity on Fermi' surface.

Let us notice, that in the limit
zero temperature ($T\to 0$) $ \mu\to \E_F=mv_F^2/2$ and
$f_F\to \Theta (1-P^2) $, where $ \Theta (x) $ -- Heaviside' function,
$$
\Theta(x)=\Big\{\begin{array}{c}
                  1, x>0, \\
                  0, x<0.
                \end{array}
$$

Besides, at $T\to 0$ $f_F^{\pm\pm}\to \Theta^{\pm\pm}$,
where
$$\Theta^{\pm\pm}=\Theta[1-(P_x\mp \hbar k/p_F)^2-P_y^2-P_z^2].
$$

Thus, the formula (2.4) for degenerate electronic plasma
it will be transformed to the following form
$$
j_{\rm long}(x,t)=
\dfrac{e^3p_F^2A^2(x,t)}{(2\pi\hbar)^3mc^2q^2}\times
$$
$$
\times \Bigg[\int
\dfrac{\Theta^{++}+\Theta^{--}-2\Theta}{(P_x-\omega/v_Tk)^2}P_xP_y^2d^3P-
\dfrac{q}{2}\int\dfrac{\Theta^{++}-\Theta^{--}}{P_x-\omega/v_Tk}P_xd^3P\Bigg],
\eqno{(3.1)}
$$

Here, in (3.1) $q=k/k_F$, $k_F=mv_F/\hbar$ is the Fermi wave number.

We notice that
$$
\dfrac{\omega}{kv_F}=\dfrac{\omega}{v_Fk_F}\cdot\dfrac{k_F}{k}=
\dfrac{\Omega}{q},\quad \Omega=\dfrac{\omega}{v_Fk_F}, \quad
\dfrac{\hbar k}{p_F}=q.
$$

Let us put in (3.1) $P_x =\tau $, $x_0 =\omega/kv_F =\Omega/q $
and we will calculate
entering into (3.1) integrals. For the first integral it is received
$$
\int
\dfrac{\Theta^{++}+\Theta^{--}-2\Theta}{(P_x-\omega/kv_F)^2}P_xP_y^2d^3P=
$$
$$
=\int\limits_{P^2<1}\Big[\dfrac{\tau+q}{(\tau+q-x_0)^2}
+\dfrac{\tau-q}{(\tau-q-x_0)^2}-\dfrac{2\tau}{(\tau-x_0)^2}\Big]
P_y^2d^3P=
$$
$$
=\dfrac{\pi}{4}\int\limits_{-1}^{1}\Big[\dfrac{\tau+q}{(\tau+q-x_0)^2}
+\dfrac{\tau-q}{(\tau-q-x_0)^2}-\dfrac{2\tau}{(\tau-x_0)^2}\Big](1-\tau^2)^2
d\tau.
$$

The second integral equals
$$
\int\dfrac{\Theta^{++}-\Theta^{--}}{P_x-\omega/v_Tk}P_xd^3P=
$$
$$
=\pi \int\limits_{-1}^{1}\Big(\dfrac{\tau+q}{\tau+q-x_0}-
\dfrac{\tau-q}{\tau-q-x_0}\Big)(1-\tau^2)d\tau.
$$

By means of two last equalities and the formula (2.5) for density
transversal electric current (3.1) we receive expression through
one-dimensional integrals
$$
j_{\rm long}(x,t)=\sigma_{l,tr}^{(2)}E_{tr}^2(x,t).
$$

Here $\sigma_{l, tr}^{(2)} $ is the quantity which it is natural
to name longitudinal - transversal (nonlinear)
conductivity of the second order,
$$
\sigma_{l,tr}^{(2)}=\Sigma_{l,tr}^{(2)}\dfrac{1}{\Omega^2q^2}J(x_0,q),
\eqno{(3.2)}
$$
where
$$
\Sigma_{l,tr}^{(2)}=-\dfrac{e^3}{32\pi^2 \hbar mv_F^2},
$$
$J(x_0,q)$ is the dimensionless coefficient of longitudinal --
transversal (nonlinear) conductivity of the second order,
$$
J(x_0,q)=\int\limits_{-1}^{1}\Bigg[\dfrac{[(\tau+q)(1-\tau^2)^2]'}
{\tau+q-x_0}+\dfrac{[(\tau-q)(1-\tau^2)^2]'}{\tau-q-x_0}-
\dfrac{2[\tau(1-\tau^2)^2]'}{\tau-x_0}\Bigg]d\tau-
$$
$$
-2q \int\limits_{-1}^{1}\Bigg[\dfrac{(\tau+q)(1-\tau^2)}{\tau+q-x_0}-
\dfrac{(\tau-q)(1-\tau^2)}{\tau-q-x_0}\Bigg]d\tau.
\eqno{(3.3)}
$$

Let us underline, that is longitudinal -- transversal conductivity is caused
that the transversal electromagnetic field leads to the longitudinal
current.

The integrals entering in (3.3), we will calculate by means of
equality (2.5).

The first integral is equal
$$
J_1=28x_0q^2+
[(x_0-q)^2-1][5(x_0-q)^2+4\tau_0(x_0-q)-1] \times
$$
$$
\times\Big[\ln\Big|\dfrac{x_0-q-1}{x_0-q+1}\Big|+
\Big\{\begin{array}{r}
         i\pi,\quad |x_0-q|<1 \\
         0,\quad |x_0-q|>1
       \end{array}\Big\}\Big]+
$$
$$
+[(x_0+q)^2-1][5(x_0+q)^2-4q(x_0+q)-1] \times
$$
$$
\times\Big[\ln\Big|\dfrac{x_0+q-1}{x_0+q+1}\Big|+
\Big\{\begin{array}{r}
         i\pi,\quad |x_0+q|<1 \\
         0,\quad |x_0+q|>1
       \end{array}\Big\}\Big]-
$$
$$
-2(x_0^2-1)(5x_0^2-1)\Big[\ln\Big|\dfrac{x_0-1}{x_0+1}\Big|+
\Big\{\begin{array}{r}
         i\pi,\quad |x_0|<1 \\
         0,\quad |x_0|>1
       \end{array}\Big\}\Big].
$$

The second integral equals
$$
J_2=4x_0q+
$$
$$
-x_0[(x_0-q)^2-1]\Big[\ln\Big|\dfrac{x_0-q-1}{x_0-q+1}\Big|+
\Big\{\begin{array}{r}i\pi,|x_0-q|<1 \\
0, |x_0-q|>1
\end{array}\Big\}\Big]+
$$
$$
+x_0[(x_0+q)^2-1]
\Big[\ln\Big|\dfrac{x_0+q-1}{x_0+q+1}\Big|+
\Big\{\begin{array}{r}
  i\pi,|x_0+q|<1 \\
  0, |x_0+q|>1
\end{array}\Big\}\Big].
$$

The found integrals we will substitute in (3.3) and we will allocate in it
the real and imaginary parts,
$$
J(x_0,q)=J_1(x_0,q)-2qJ_2(x_0,q)=R(x_0,q)+i\pi S(x_0,q),
\eqno{(3.4)}
$$
believing, that $x_0=\Omega/q$.

We receive that
$$
R(\Omega,q)=20\Omega q+\Big[\Big(\dfrac{\Omega}{q}-q\Big)^2-1\Big]
\Big[5\dfrac{\Omega^2}{q^2}-4\Omega+q^2-1\Big] \times
$$
$$
\times\ln\Big|\dfrac{\Omega-q^2-q}{\Omega-q^2+q}\Big|+
\Big[\Big(\dfrac{\Omega}{q}+q\Big)^2-1\Big]
\Big[5\dfrac{\Omega^2}{q^2}+4\Omega+q^2-1\Big] \times
$$
$$
\times\ln\Big|\dfrac{\Omega+q^2-q}{\Omega+q^2+q}\Big|-2
\Big(\dfrac{\Omega^2}{q^2}-1\Big)\Big(5\dfrac{\Omega^2}{q^2}-1\Big)
\ln\Big|\dfrac{\Omega-q}{\Omega+q}\Big|,
$$
and

$$
S(\Omega,q)=\Big[\Big(\dfrac{\Omega}{q}-q\Big)^2-1\Big]
\Big[5\dfrac{\Omega^2}{q^2}-4\Omega+q^2-1\Big] \Big\{
\begin{array}{l}1,|\Omega-q^2|<|q| \\0,|\Omega-q^2|>|q|\end{array}\Big\} +
$$
$$
+\Big[\Big(\dfrac{\Omega}{q}+q\Big)^2-1\Big]
\Big[5\dfrac{\Omega^2}{q^2}+4\Omega+q^2-1\Big] \Big\{\begin{array}{l}
  1,|\Omega+q^2|<|q| \\0,|\Omega+q^2|>|q|\end{array}\Big\}-
$$
$$
-2\Big(\dfrac{\Omega^2}{q^2}-1\Big)\Big(5\dfrac{\Omega^2}{q^2}-1\Big)
\Big\{\begin{array}{l}1,|\Omega|<|q| \\
0,|\Omega|>|q|\end{array}\Big\}.
$$

According to (3.2) and (3.4) for longitudinal -- transversal
conductivity we receive following expression
$$
\sigma_{l,tr}^{(2)}=\Sigma_{l,tr}^{(2)}
\dfrac{R(\Omega,q)+i\pi S(\Omega,q)}{\Omega^2q^2}.
\eqno{(3.5)}
$$

Let us underline, that in expression (3.5) are allocated Kohn's
singu\-la\-ri\-ties of the form $X\ln X $.
It means, that expression (3.5) has no
singularities in zero of logarithms, i.e. in those points
$ (\Omega_0, q_0) $, in which $X (\Omega_0, q_0) =0$.

\begin{center}
  \bf 4. Conlusions
\end{center}

On Figs. 1 -- 6 we will present behaviour of coefficients
$R $ and $S $ depending on dimensionless frequency of
oscillations of the vector poten\-tial $ \Omega $ and dimensionless
wave number $q $.

On Fig. 1 and 2 we will represent behaviour of coefficient
$R $ in dependence from frequency $ \Omega $ at various values
$q $. From these plots it is visible, as at small values $q $ and at
values $q $, comparable with unit, coefficient $R $,
proportional to the real part of the generated longitudinal
current, has at first a minimum, and then a maximum, and a minimum
lays near to a point $ \Omega=q $.

On Fig. 3 and 4 we will represent behaviour of  coefficient
$R $ in dependence from wave number $q $ at various values of
frequency $ \Omega $.
In both cases at small values of frequency and at values
frequencies near to unit the coefficient $R $ has at first a maximum, and
then the minimum, and a minimum is found out nearby
considered value
$$
\Omega=q=0.10,\quad 0.11,\quad 0.12.
$$

On Fig. 5 and 6 the behaviour of coefficient $S $ as functions
wave number at various small values of oscillations frequency
is represented. At small values of frequency of oscillations
coefficient $S $ has a minimum. With increase $ \Omega $ the
coefficient $S $ can to have two minima and one maximum.
This maximum vanishes with growth $ \Omega $.

From Fig. 7 it is visible, that at $ \Omega=1$ function $S=S (\Omega, q) $
has in the point $q=1$ a local maximum, and near to this point
at the left and to the right of it has two more local minima. At
$ \Omega=2$ function $S=S(2, q) $ in a point $ \Omega=2$ has the local
maximum, and at the left and to the right of it has two more local
minimum.

On Fig. 8 we observe the similar similar situation for three
curves $S=S (0.7, q), S=S (1, q) $ and $S=S (1.3, q) $.

In the present work the account of nonlinear character of
interaction electromagnetic field with quantum plasma is
considered. It has appeared, that
the account of nonlinearity of an electromagnetic field finds out
generating of an electric current, orthogonal to a direction
fields.

\clearpage

\begin{figure}[t]\center
\includegraphics[width=16.0cm, height=10cm]{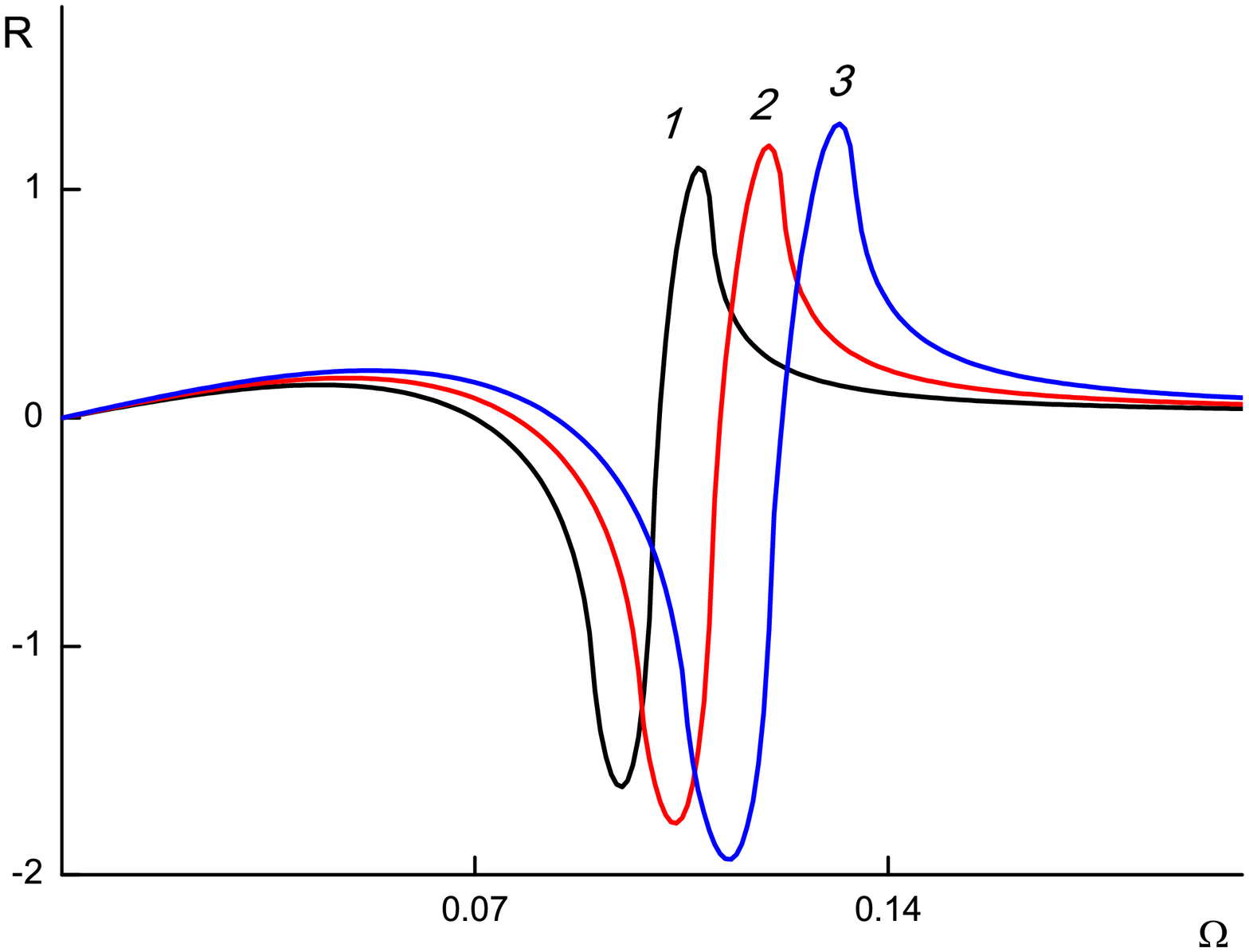}
{{ Fig. 1.  Real part of coefficient $R(\Omega,q)$.
Curves 1,2,3 correspond to values of dimensionless wave number
$q=0.1,0.11,0.12$.}}
\includegraphics[width=16.0cm, height=10cm]{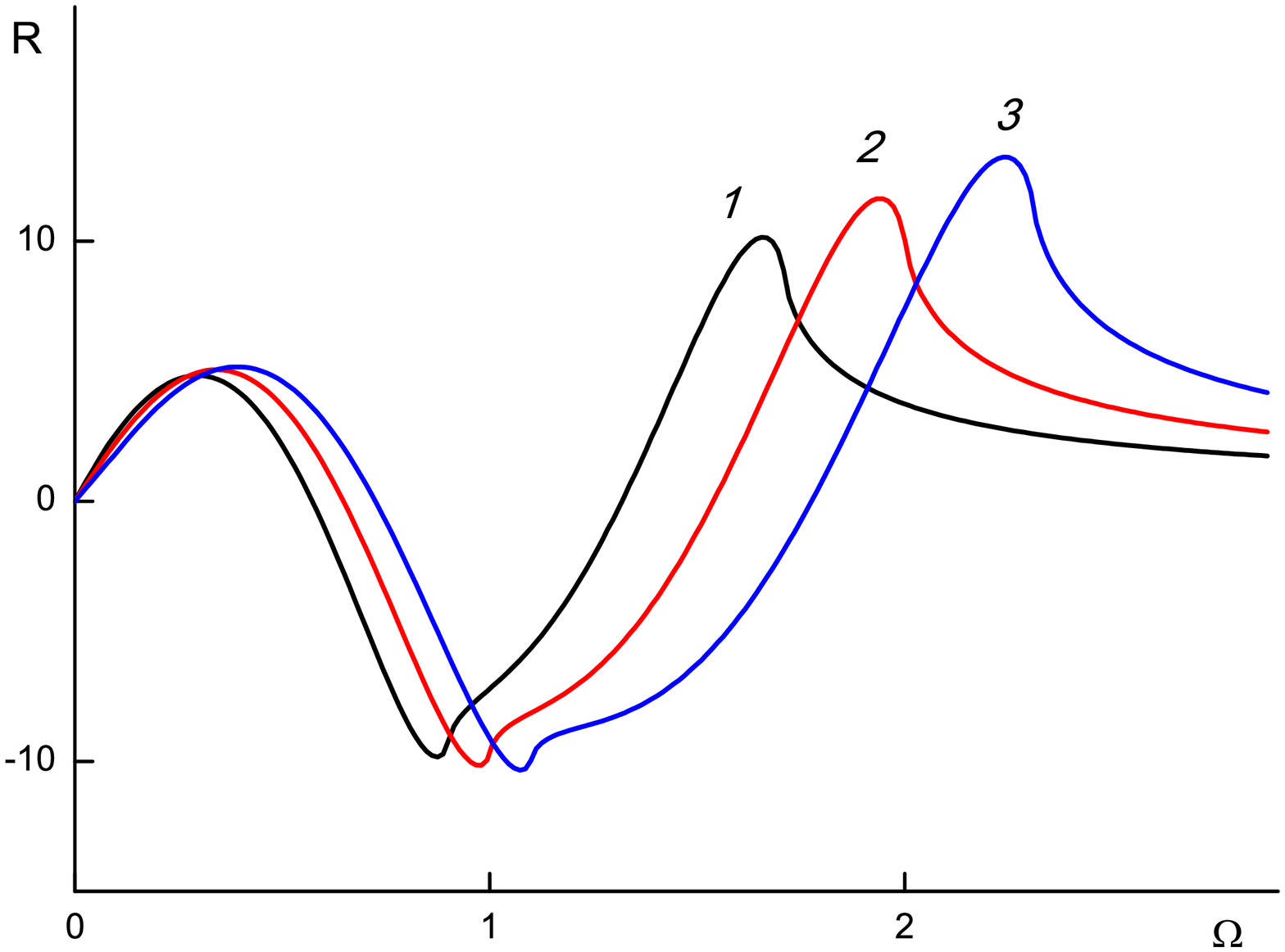}
{{Fig. 2. Real part of coefficient $R(\Omega,q)$.
Curves 1,2,3 correspond to values of dimensionless wave number
$q=0.9,1.0,1.1$.}}
\end{figure}

\begin{figure}[th]\center
\includegraphics[width=16.0cm, height=10cm]{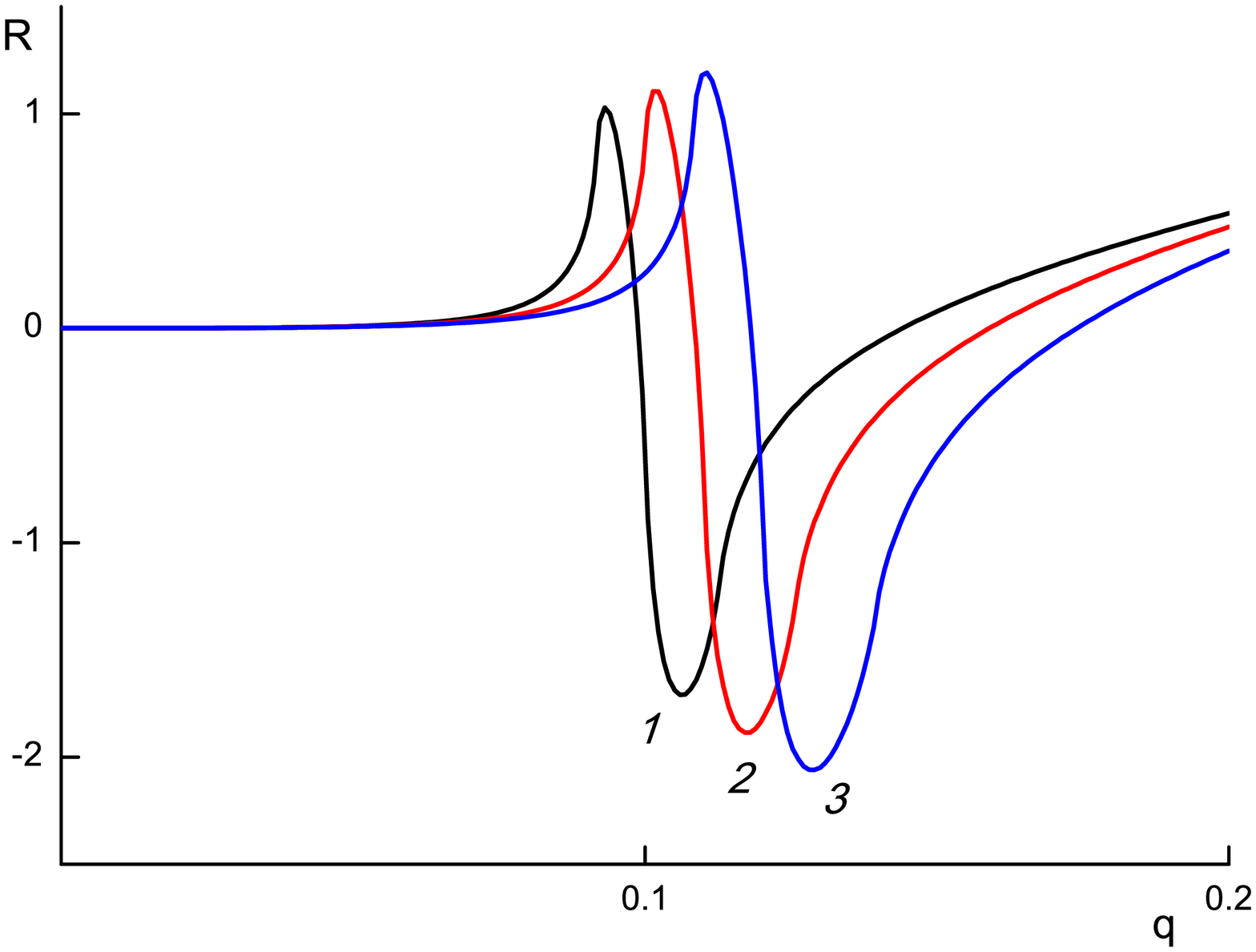}
{{ Fig. 3. Real part of coefficient $R(\Omega,q)$.
Curves 1,2,3 correspond to values of dimensionless frequency
$\Omega=0.1,0.11,0.12$.}}
\includegraphics[width=16.0cm, height=10cm]{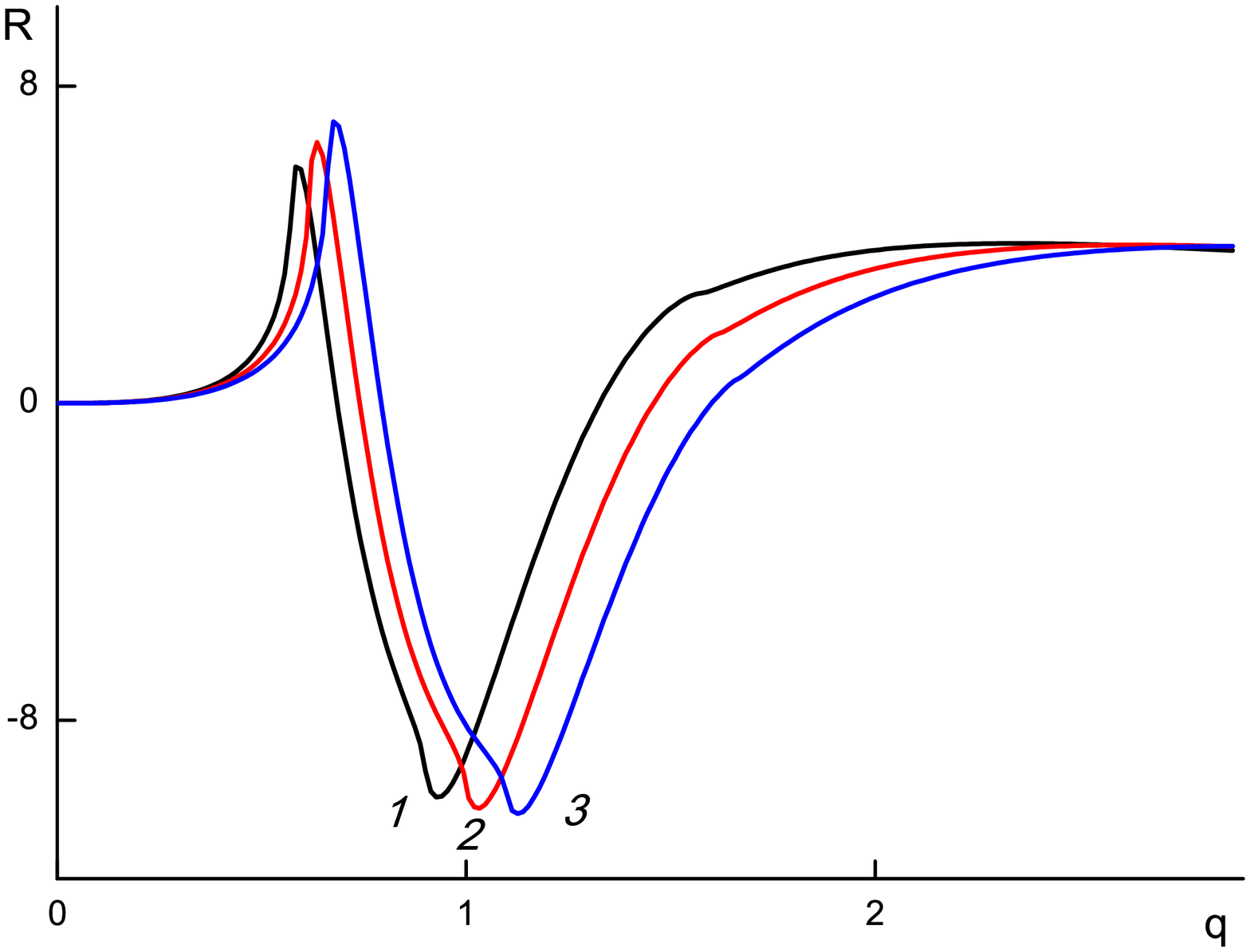}
{{ Fig. 4. Real part of coefficient $R(\Omega,q)$.
Curves 1,2,3 correspond to values of dimensionless frequency
$\Omega=0.7,1.0,1.3$.}}
\end{figure}

\begin{figure}[thp]\center
\includegraphics[width=16.0cm, height=10cm]{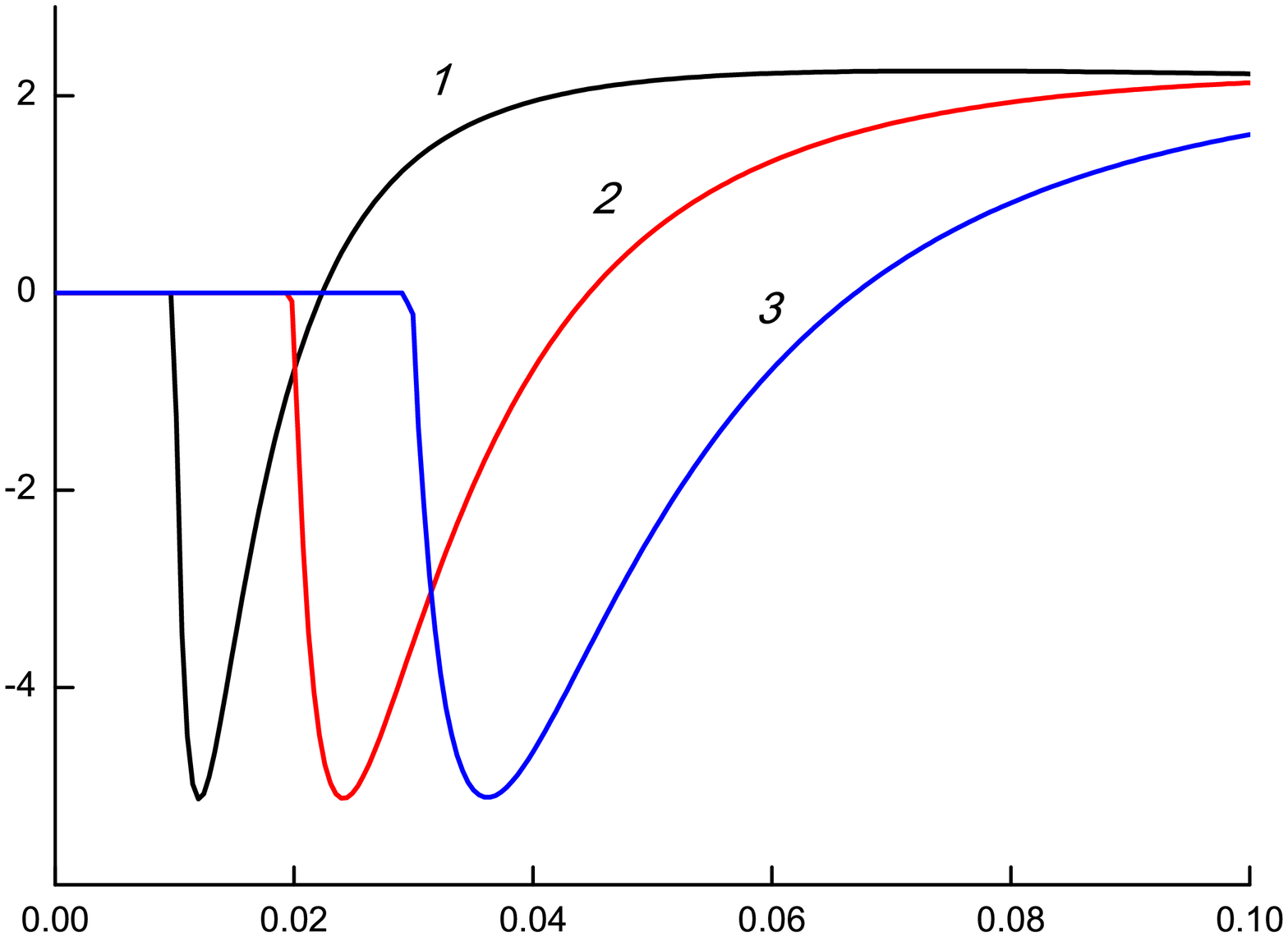}
{{ Fig. 5.  Imaginare part of coefficient $S(\Omega,q)$.
Curves 1,2,3 correspond to values of dimensionless frequency
$\Omega=0.01,0.02,0.03$.}}
\includegraphics[width=16.0cm, height=10cm]{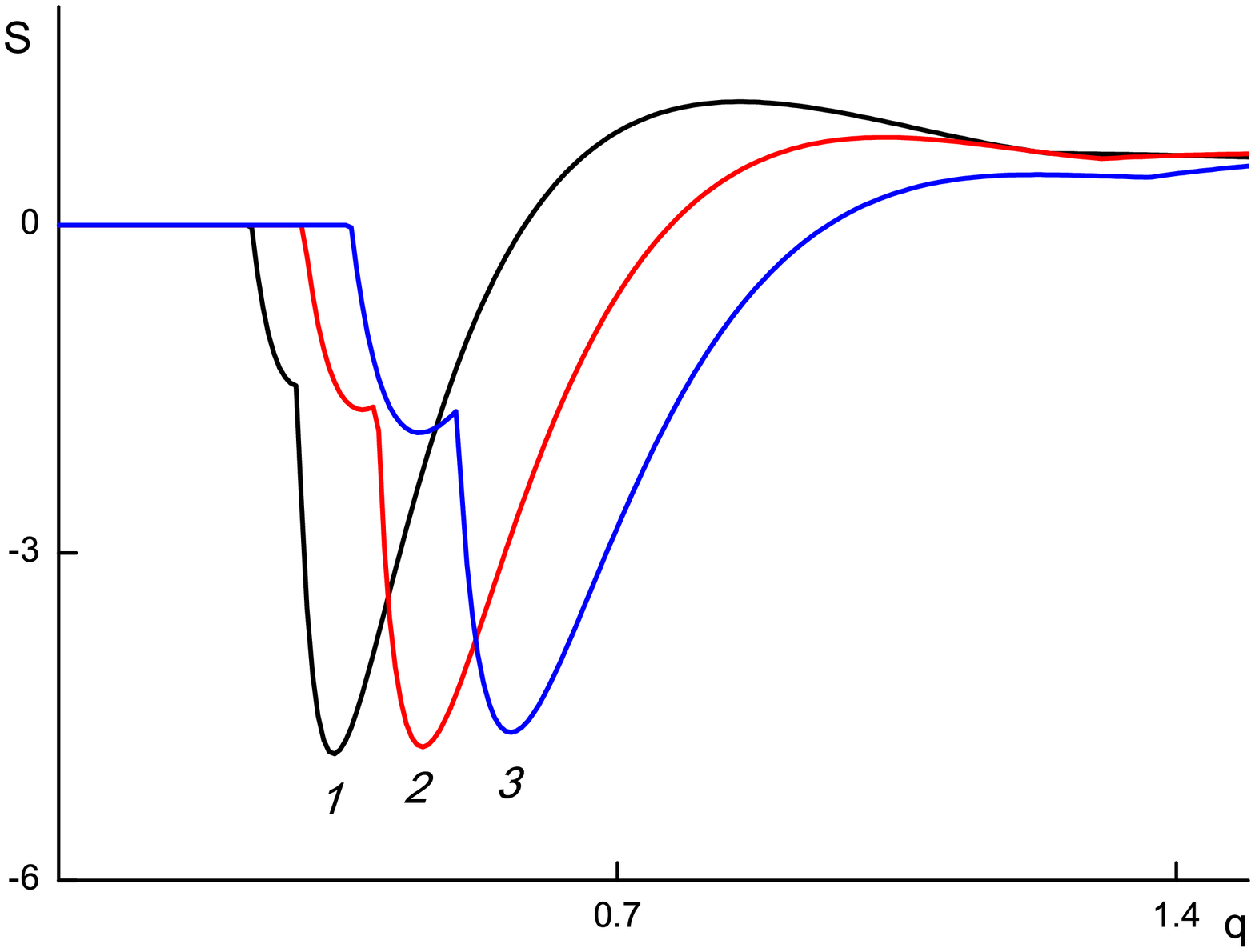}
{{ Fig. 6. Imaginare part of coefficient $S(\Omega,q)$.
Curves 1,2,3 correspond to values of dimensionless frequency
$\Omega=0.3,0.4,0.5$.}}
\end{figure}

\begin{figure}[thp]\center
\includegraphics[width=16.0cm, height=10cm]{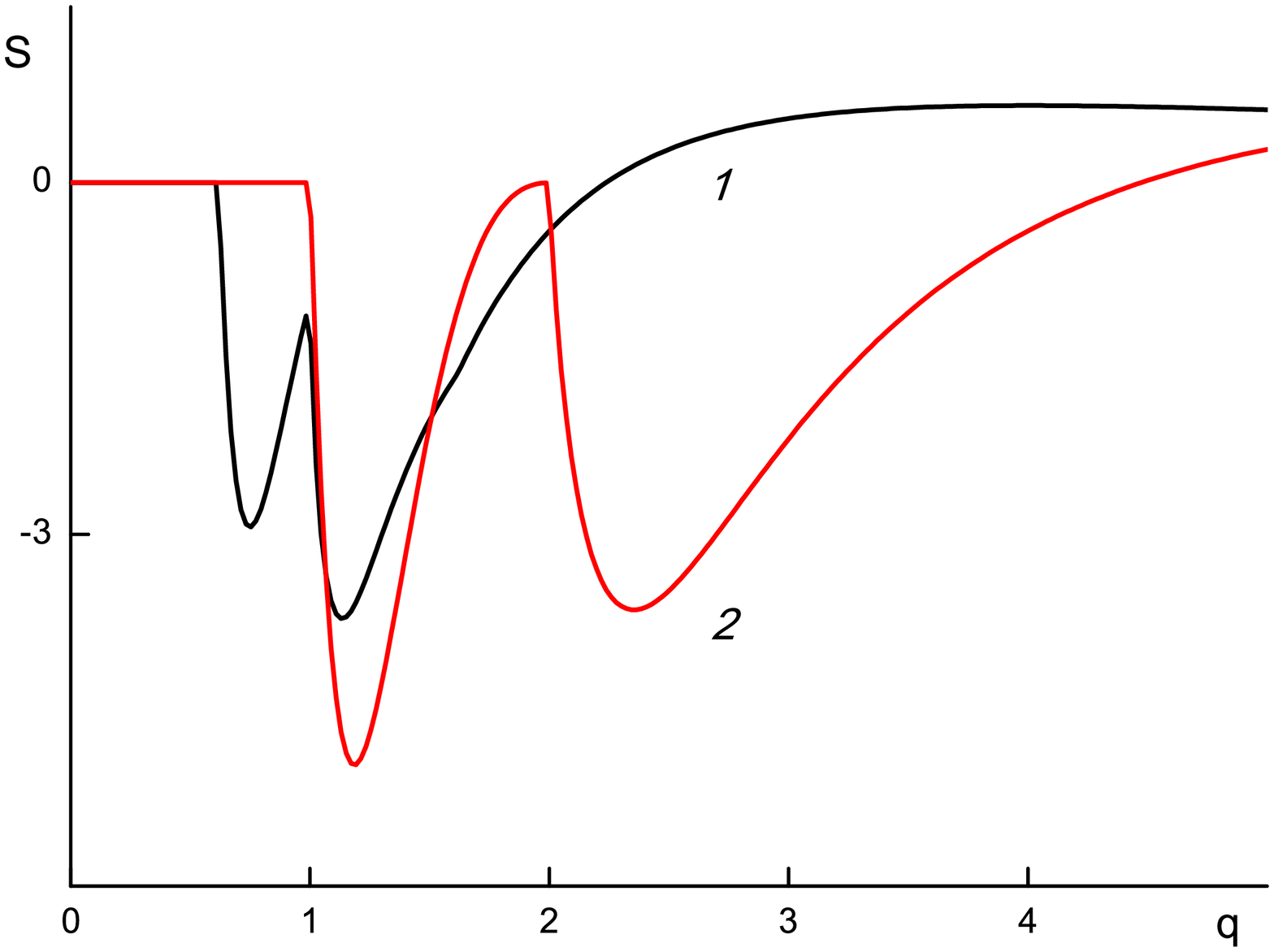}
{{ Fig. 7.  Imaginare part of coefficient $S(\Omega,q)$.
Curves 1 and 2 correspond to values of dimensionless frequency
$\Omega=1$ and $\Omega=2$.}}
\includegraphics[width=16.0cm, height=10cm]{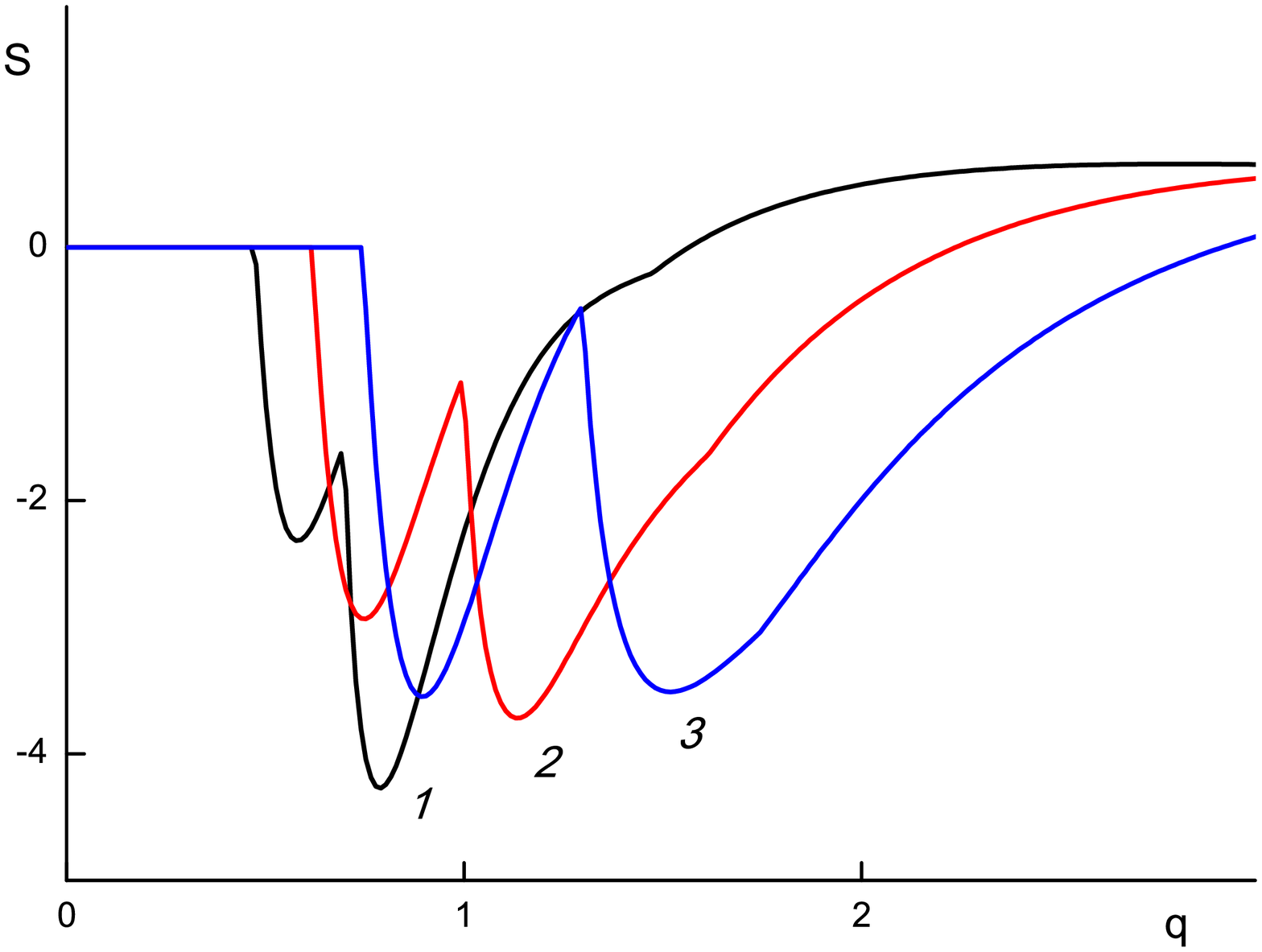}
{{ Fig. 8. Imaginare part of coefficient $S(\Omega,q)$.
Curves 1,2,3 correspond to values of dimensionless frequency
$\Omega=0.7,1.0,1.3$.}}
\end{figure}

\end{document}